# TWO CHALLENGES OF STEALTHY HYPERVISORS DETECTION: TIME CHEATING AND DATA FLUCTUATIONS

Igor Korkin

National Research Nuclear University Moscow Engineering & Physics Institute (NRNU MEPhI)
Department of Cryptology and Discrete Mathematics
Moscow, 115409, Russia
igor.korkin@gmail.com

## ABSTRACT

Hardware virtualization technologies play a significant role in cyber security. On the one hand these technologies enhance security levels, by designing a trusted operating system. On the other hand these technologies can be taken up into modern malware which is rather hard to detect. None of the existing methods is able to efficiently detect a hypervisor in the face of countermeasures such as time cheating, temporary self-uninstalling, memory hiding etc. New hypervisor detection methods which will be described in this paper can detect a hypervisor under these countermeasures and even count several nested ones. These novel approaches rely on the new statistical analysis of time discrepancies by examination of a set of instructions, which are unconditionally intercepted by a hypervisor. Reliability was achieved through the comprehensive analysis of the collected data despite its fluctuation. These offered methods were comprehensively assessed in both Intel and AMD CPUs.

## Contents

# 1. INTRODUCTION

Nowadays successful malware detection is becoming increasingly important, because malware cyber-attacks can result in financial, reputational, process and other losses. We can overcome these risks only through anticipatory development of advanced cyber security solutions.

Intel and AMD have released more advanced CPUs with hardware virtualization support, which runs code directly on top of the physical hardware. This privileged code is named Virtual Machine Monitor (VMM), bare-metal hypervisor or just "hypervisor". A hypervisor with a secure system monitor functions allows us to run multiple OSes at the same time in one PC, (see Figure 1). As a result this architecture maximizes the hardware utilization and reduces the costs of operation. This is an obvious advantage of hardware virtualization based hypervisors (Derock, 2009; Barrett, & Kipper, 2010). At present more than a billion processors with this technology are installed in workstations as well as in cloud computing servers on the Internet.

However, at the same time hardware virtualization technology increases vulnerability of systems, seeing that rootkit hypervisor with backdoor functionality can be planted in the PC (Ben-Yehuda, 2013). This type of rootkits is also knows as Hardware-based Virtual Machine Rootkit (HVM rootkit).

The cyber security community faces the challenge of hypervisor detection. Presently there is no built-in tool to detect a hypervisor reliably. Of course we can check basic things: CR4.VMXE bit in Intel case (Intel, 2014) or EFER.SVME bit in AMD case (AMD, 2013), but a hypervisor can hide its original value. Moreover, it is impossible to block, stop or unload a hypervisor by using existing known cyber security tools, resides on virtualized OS level.

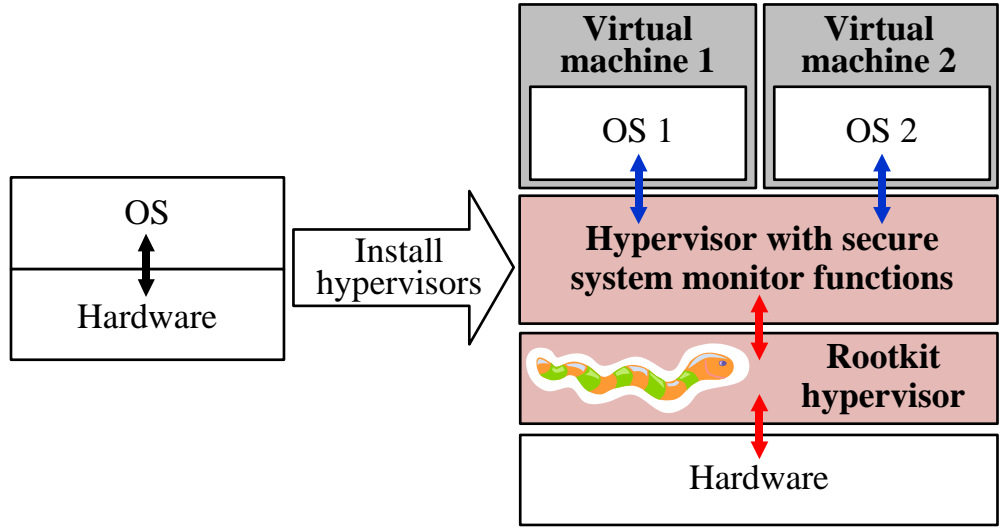

Figure 1 PC without Hypervisor and under Control of the Two Nested Hypervisors: a Legitimate one and Rootkit

The difficulties of this challenge arise from the following causes. Firstly hypervisors can use a wide variety of different techniques to prevent detection. Secondly, it is possible to run several nested hypervisors. Thirdly, a hypervisor can be installed via a driver or boot records as well as via BIOS (Kovah, Kallenberg, Butterworth, & Cornwell, 2014) or UEFI (Bulygin, Loucaides, Furtak, Bazhaniuk, & Matrosov, 2014), which makes the deleting of a hypervisor rather difficult.

Utin (2014) analyzed the possibility of BIOS-based hypervisor threat. The author's ideas are based on the suspicious hypervisor (Russian Ghost) whose detection is simple, because it does not apply any countermeasures.

Despite the fact that hardware virtualization is not new and involves a world-wide community of researchers, the development of effective hypervisor detection methods has so far been without success.

The goal of this paper is to tackle this issue. This article presents new detection methods which are based on the difference between the instruction execution time (IET) both, with a hypervisor and without it. We applied a set of specific instructions which cause VM-exits unconditionally or are trapped by a hypervisor. As a result, IET takes significantly more time with a hypervisor than without any hypervisor.

This time discrepancy is commonly used to detect hypervisors. However, detection by time is possible only if a hypervisor is not hiding itself via timestamp cheating (Fritsch, 2008; Garfinkel, Adams, Warfield, & Franklin, 2007) or via a temporary self-uninstalling hypervisor – the Blue Chicken technique (Rutkowska, & Tereshkin, 2007). Under these conditions the hypervisor detection methods based on time discrepancies will not work. Therefore, a totally new hypervisor detection approach, which is resilient to countermeasures, is needed.

In a nutshell the proposed methods consider the IET as a random variable, whose properties depend on hypervisor presence. That is why by applying probabilistic and statistical methods to IET, it may be possible to detect a hypervisor.

Our detection methods have improved on the current time-based detection method, which uses unconditionally intercepted instructions. Unlike the original method our approach is able to detect any stealthy hypervisor, which has applied countermeasures: time-cheating, temporary self-uninstalling etc. This is a distinct advantage of these new methods.

The remainder of the paper is organized as follows. Section 2 is devoted to the analysis of the most popular software and hardware hypervisor detection approaches. The analysis will be given in the case of a hypervisor using countermeasures to prevent its detection, such as time cheating, temporary self-uninstalling, preventing memory dump acquisition etc.

Section 3 contains the processor behavior analysis in the three cases without a hypervisor, with one and several nested hypervisors. Analysis has discovered new useful *statistics for the IET*, which can reveal hypervisors.

In section 4 the experimental results of statistics examination are presented. The positive results of these checks make it possible to analyze IET as a random variable. As a result this allows us to use threshold values of statistics to detect each hypervisor. This approach works well under the countermeasures and fluctuations of measured time durations. The present author's threshold generated methods and hypervisor detection approaches and their analysis are briefly presented.

Section 5 contains the main conclusions and further research directions.

## 2. RELATED WORK

Nowadays there is no hypervisor detection build-in tool for Intel. The built-in tool for AMD CPU is vulnerable to hypervisor countermeasures. Therefore researchers are working hard to solve this challenge. This paper gives a classification and analysis of all publicly available hypervisor detection methods and approaches.

The history of hypervisor detection started in 2007 after the first hypervisor rootkit "Blue Pill" was presented by Rutkowska (2006). "Blue Pill" is a Windows based driver for AMD CPU. At the same time Dai Zovi (2006) released "Vitriol" – a similar hypervisor for MAC OS and Intel CPU.

The comparative analysis of these two hypervisors was presented by Fannon (2014). "Blue Pill" and "Vitriol" became high-profile tools in information security sphere and motivated the creation a lot of different approaches to hypervisor detection. Their classification is given in Figure 2. We can classify these into four categories: signature-based, behavior-based, detection based on the trusted hypervisor and approaches which use time analysis. Signature-based detection uses memory scanning of hypervisors' patterns. The latter three sections are based on interaction with a hypervisor.

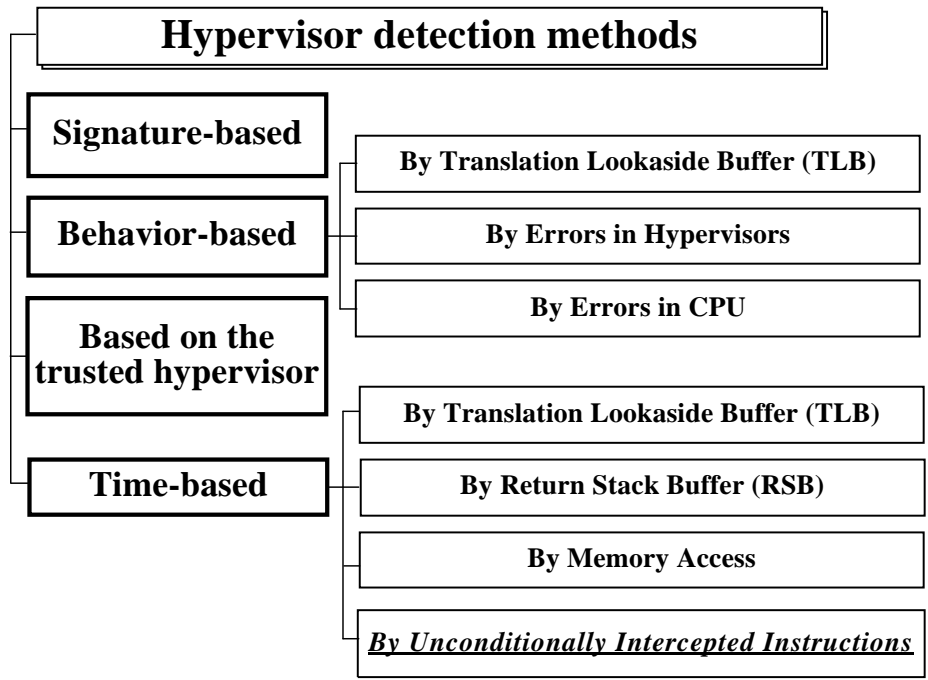

Figure 2 Hypervisor Detection Methods Classification

### 2.1. Signature-Based Detection

After a hypervisor has been loaded into memory its dispatcher (VMM handler) and Virtual Machine Control Structure (VMCS in Intel case) will be located in memory. The hypervisor can be detected by signature analysis of the physical memory (Bulygin & Samyde, 2008; Desnos, Filiol, & Lefou, 2011; Medley, 2007).

This approach consists of two stages: memory dump and its inspection, both of which are not resilient to the hypervisor's countermeasures. Analysis shows that software based memory dump approaches are vulnerable whereas the hardware ones are only applicable under laboratory conditions (Korkin & Nesterov, 2014).

Let us analyze how resistant the current hypervisor's signatures are.

Thus Fritsch (2008) proposed to detect "Blue Pill" hypervisor by searching "BLPB", "BLUE" and "BLUP" strings in a memory dump. However, in common cases such strings will be unknown to analysts.

The Actaeon system (Graziano, Lanzi, & Balzarotti, 2013) is based on searching for VMCS fragments. However, this method can sometimes fail. For example, hypervisor can allocate in memory 100 structures to hamper detection. These VMCSes are similar to original VMCS. After that the Actaeon system may reveal many false VMCSes so separation between the original one and the rest will require a considerable amount of manual work.

As a result, signature-based detection is ineffective for resistant hypervisors.

### 2.2. Behavior-Based Detection

Behavior-based detection relies on the system activity differences in the two cases, with and without a hypervisor. There are three behavior-based detection methods: TLB-based detection and methods based on errors in hypervisors and errors in CPUs.

**2.2.1. TLB-Based Detection.** It is possible to apply the Translation Lookaside Buffer (TLB) which is a memory cache used to speed address translation to detect a hypervisor (Desnos et al., 2011; Fritsch, 2008; Morabito, 2012; Wailly, 2014).

TLB includes a set of recently accessed virtual and corresponding physical addresses. Every time OS accesses memory a corresponding TLB entry is searched for. If the requested virtual address is present in the TLB, the retrieved physical address will be used to access memory. In the other case the longtime search with the help of Page Directory will occur. This peculiarity will be discussed later in Section 2.4.1.

It is known that VM-exit leads to flushing of TLB when a hypervisor is present. Otherwise without a hypervisor such clearance does not occur. This is why hypervisor detection reduces to checking TLB content, which can be made in several ways, for example by modifying page table entry (Myers & Youndt, 2007).

However, TLB-based detection does not work on AMD CPUs and new Intel CPUs. The new supplementary TLB fields "ASID" and "PCID" do not let VM-exit flush TLB.

**2.2.2. Detection Based on Bugs in CPU.** A hypervisor can be detected with the help of bugs in certain CPU models. In these CPUs the results of some instructions depend on whether or not a hypervisor is present.

The "Erratum 140" in AMD CPU is based on using results of "RDMSR 10h". The original value of the Time Stamp Counter (TSC) is returned by "RDMSR 10h" while "RDTSC" gets the sum of TSC value and VMCS.TSC_OFFSET value (AMD, 2011).

Another bug "VMSAVE 0x67" freezes the system. The execution of the VMSAVE instruction with 0x67 prefix stops virtualization system. Without a hypervisor this error does not occur (Barbosa, 2007).

These detection methods are applicable only for outdated CPUs and require non trivial adaptation to new CPUs.

**2.2.3. Detection Based on Bugs in Hypervisors.** There are software hypervisor bugs similar to hardware bugs in CPU.

Microsoft published their paper "Hypervisor Top-Level Functional Specification", which describes how to detect a hypervisor and get "Hypervisor Vendor ID Signature", by using CPUID (O'Neill, 2010; Microsoft, 2013). Spoofing attack is likely to occur, when a hypervisor can replace data, trapped by CPUID execution.

"Blue Pill" hypervisor has a built-in control interface, which uses "Bpknock" hypercalls (BluePillStudy, 2010; Fritsch, 2008). Calling CPUID with EAX=0xbabecafe changes EAX to 0x69696969, if "Blue Pill" is present. Otherwise such a change does not occur. Due to the hypervisor's built-in control interface it is possible not only to detect, but also unload a hypervisor (Gabris, 2009).

A hypervisor can also be detected by reading debugging messages. For example, a developer or hacker might have forgotten to remove DbgPrint calls, which can disclose a hypervisor's activity.

These approaches can reveal only well-known hypervisors, which do not take countermeasures.

### 2.3. Detection Based on the Trusted Hypervisor

A hypervisor which is loaded first can control and block activity of hypervisors which are loaded later. This detection method was used in "McAfee DeepSAFE" (McAfee, 2012), "Hypersight Rootkit Detector" (North Security Labs, 2011), "Symantec Endpoint Protection" (Korkin, 2012), and it has also been mentioned in papers (Park, 2013; Wang & Jiang, 2010).

This detection approach is vulnerable to "Man-In-The-Middle" (MITM) attack, in which an illegitimate hypervisor can gain control first and compromise a legitimate one, which was loaded later on. TPM-based attestation of hypervisor can avoid this attack, although TMP mechanism is vulnerable too (Berger et al., 2006; Brossard & Demetrescu, 2012; Wojtczuk, & Rutkowska, 2009; Wojtczuk, Rutkowska, & Tereshkin, 2009).

MITM attack can be also prevented by loading hypervisor from BIOS as well as by applying Trusted Startup Hardware Module (Accord, 2010). However, due to the difficulty of porting this detection method, it is applicable only to labs.

### 2.4. Time-Based Detection

Time-based detection uses the measuring of time duration of specific operations or profiling of its execution time. When a hypervisor is present the execution of such operations is intercepted by the hypervisor. As a result, their duration will be longer than without a hypervisor.

Four time-based methods can be mentioned: TLB- and RSB-based detection, detection based on memory access and detection by unconditionally intercepted instructions. Let us focus on these methods applicable in the situation where a hypervisor prevents its detection by time cheating and temporary self-uninstalling.

**2.4.1. TLB-Based Detection.** As it was mentioned before in Section 2.2.1, the TLB flushes every time VM-exit occurs. After that, the longtime fill will happen. It is possible to use this fact to detect hypervisor as follows (Ramos, 2009; Rutkowska, 2007):

1. Read the content of a specific memory address.
2. Repeat step 1 and measure its duration. In this case the TLB entry, which was added on step 1, will be used.
3. Execute unconditionally intercepted instruction (forcing a VM-exit).
4. Repeatedly carry out step 2.
5. Make a conclusion about the hypervisor presence by comparing the results of steps 2 and 4.

This approach does not work if the hypervisor uses time cheating, because there is no significant difference between these two steps. This approach has the same disadvantages as in Section 2.2.1.

**2.4.2. RSB-Based Detection.** Another detection method is based on Return Stack Buffer (RSB), which increases computer performance. RSB content as well as TLB suffers changes when VM-exit occurs, but unlike TLB, RSB includes addresses of RET instructions.

Applying RSB to hypervisor detection was described by Bulygin (2008) and later by Fritsch (2008) and Athreya (2010). After 16 nested functions calls, RSB will consist of 16 corresponding return addresses. The idea of the detection lies in an attempt to fill the RSB buffer, call VM-exit, for example by calling an unconditionally intercepted instruction, measure an execution time of these 16 functions. If a hypervisor is present, it intercepts VM-exit and replaces a part of RSB entries. As a result the whole duration will be longer than without a hypervisor.

This method is vulnerable to the hypervisor's countermeasures, for example if a hypervisor dispatcher has no sub-functions it is also vulnerable to time cheating attack (Athreya, 2010).

### 2.4.3. Detection Based on Memory Access.
A hypervisor can prevent its signature detection by controlling memory access (section 2.1.), which increases the duration of memory access and can be applied to hypervisor detection (Fisher-Ogden, 2006; Fritsch, 2008).

By walking successively through memory we measure each time the duration of memory page access. The memory region with excessive access duration is the stealth memory region. This region can consist of hypervisor dispatcher and corresponding structures.

However, this method works only if the hypervisor does not use time cheating for self-protection.

### 2.4.4. Detection by Unconditionally Intercepted Instructions.
It is known that the duration of execution of unconditionally intercepted instructions increases after any hypervisor has been loaded in the system. We can detect hypervisor presence by comparing time duration with some threshold values (Athreya, 2010; Lakshminarayanan, Patel, Robinson, & Soulami, 2012).

Hardware virtualization for Intel CPU includes a set of unconditionally intercepted instructions, e. g. CPUID (Intel, 2014), for AMD CPU case we can use RDMSR (Morabito, 2012), which has to be triggered by a hypervisor. The authors also suggest measuring a HDD access time, RAM access time or duration of cryptographic computation (Kyte, Zavarsky, Lindskog, & Ruhl, 2012; Pek, & Buttyan, 2014). But such events can only be revealed by specialized hypervisors and does not work in ordinary cases.

This detection approach is vulnerable to "Blue Chicken" technique and time cheating (Rutkowska, & Tereshkin, 2008). Nevertheless, this approach appears to be the most attractive because of its usability and portability. This approach is also universal, as a hypervisor will always spend time on VM-exits (VM-entries), and this time needs to be hidden. Because of these advantages this approach was chosen and was significantly improved.

## 2.5. Analysis of Counters to Measure Instruction Execution Time

Instruction execution time (IET) or its duration is the main scope of this research, so let us classify and analyze the capabilities of the computer counters, which can be applied to measure, e.g. the execution time of ten CPUID instructions.

Counters can be classified as software and hardware ones. Hardware counters use device capabilities and may be further classified as local and remote ones.

The software counter (or SMP counter) is based on simultaneous work of two loops (Desnos et al., 2011; Jian, Huaimin, Shize, & Bo, 2010; Morabito, 2012), which are running on different CPU cores. The first thread increments the control variable, while the second one executes the unconditionally intercepted instruction in the loop, for example 1000 times. The conclusion about hypervisor presence is made by comparing the results of a control variable with the threshold value. One paper (Li, Zhu, Zhou, & Wang, 2011) describes how to prevent this approach by applying memory modification, which contains the control variable.

To measure IET we can use the following hardware counters TSC, RTC, ACPI Timer, APIC Timer, HPET, PIT, local device counters, e.g. GPU timer, and NTP-based clock. Our analysis shows that all these counters apart from TSC and SMP have low-resolution and cannot be used in ordinary cases. SMP counting requires no less than two CPU cores and can be cheated. The best choice to measure the IET is TSC because of its accuracy and high-resolution. TSC also works on all CPUs. To eliminate the influence of other running programs on IET, we can use TSC on the highest IRQL and set the affinity of the measuring code with one of the CPU cores.

The important advantage of TSC is the possibility to cheat on it easily, so we can simulate a stealthy hypervisor and test our detection approach in a real case.

## 2.6. Conclusion

The above analysis shows that the existing approaches and hypervisor detection tools have the following drawbacks:

1. Signature-based approaches are vulnerable to hypervisor countermeasures. Only Actaeon project can detect nested hypervisors, but it can also be compromised.
2. Behavior-based detection methods do not reveal new hypervisors and do not work on new CPUs.
3. Trusted hypervisor-based approach is susceptible to MITM attack.
4. Time-based detection approaches are vulnerable to time cheating and Blue Chicken technique.

Detection by unconditionally intercepted instructions is highly attractive, because it relies on a generally applicable technique. By improving data acquisition and processing, we can overcome the drawbacks of this method.

## 3. THEORETICAL PRINCIPLES FOR ENHANCEMENT OF TIME-BASED DETECTION

Detection by unconditionally intercepted instructions works well only if a hypervisor does not apply countermeasures: time cheating and temporary self-uninstalling. In this section the enhancement of this method is described.

Our prerequisites are based on specific features of IET. One of them is the relation between the average IET and presence of a hypervisor. Another well-known one is a *random nature of IET*, but it is still unclear how to use it in practice.

To grapple with this gap, let us look at the switching schemes between different CPU operating modes, which occur after OS is loaded.

We demonstrate and analyze what actually happens when a set of CPUID instructions are being executed in three cases: when the hypervisor is present, not present and when several nested ones are present.

Further we will focus on two IET characteristics: variance of IET array and IET array layering.

According to some papers (Duflot, Etiemble, & Grumelard, 2006; Embleton, Sparks, & Zou, 2008; Zmudzinski, 2009) without a hypervisor a CPU can operate in one of the two modes: either in the Protected Mode (P-mode) or System Management Mode (S-mode), which is depicted on Figure 3, *a*. System Management Interrupt (SMI) switches from the P- to S-mode, CPU leaves S-mode and returns to the previous mode by using RSM instruction.

We can conclude that CPU is a stochastic system with random transitions between states, because of a random nature of SMI. Therefore IET is a random value determined by the number of SMI.

After the hypervisor is loaded the CPU can switch between the three modes. As in the previous case the P- and S- modes are present but an additional VMX root mode (V-mode) is added, so the P-mode is named as VMX non root mode (Intel, 2014). The P-mode is accepted as the main one, S-mode is duplicated for better clarity, see Figure 3, *b*. Execution of each CPUID instruction in P-mode always leads to switching to the V-mode (VM-exit), and after execution it switches back to the P-mode. Switching to the S-mode might occur either from P-mode or from V-mode.

Similarly to the previous case we may assume that CPU works as a stochastic system, but switching to the V-mode enhances its random nature. As a result switching increases the average value of IET as well as the variability of IET.

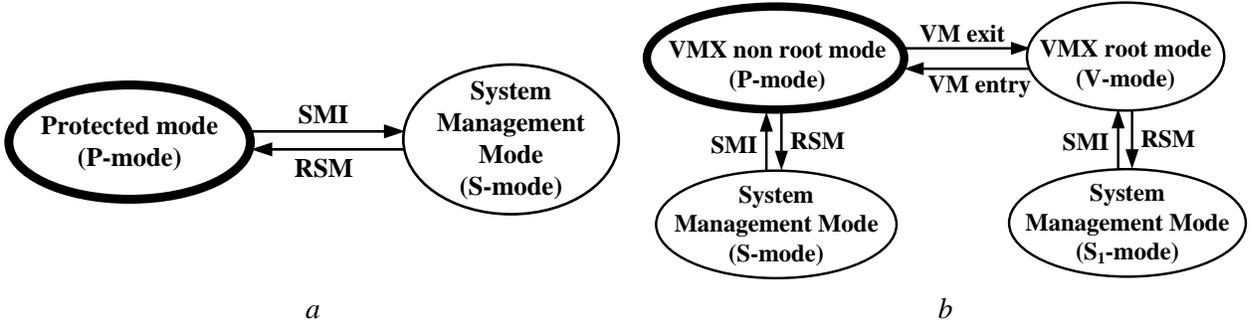

Figure 3 Switching between Modes in Two Cases: (*a*) without a Hypervisor, (*b*) with One Hypervisor

CPU works in a similar way in cases when several hypervisors are present (Ben-Yehuda et al., 2010). CPU can also switch between three modes, but the situation will be different because of several hypervisors dispatchers, see Figure 4.

In this case execution of each CPUID instructions in P-mode always leads to switching to the V-mode, and further, each hypervisor's dispatcher is successively called beginning from dispatcher #1 to dispatcher #2 etc. to dispatcher #*e* and backwards. Finally execution will switch to P-mode. S-mode can gain control at any point. Now, CPU also works as a stochastic system, but participation of several nested dispatchers significantly lengthens the time of execution and increases IET variability. These schemes allow us to discover that the root of randomness of IET is actually the randomness of SMI.

Suppose that probability or frequency of SMI is a constant. After a hypervisor is loaded, due to the increased IET the number of SMI is increased as well. That is why the variance of IET will increase after a hypervisor is loaded and this fact can be used for detection. During the execution of a set of CPUID the number of SMI is limited. If we repeat measuring of IET in a loop we can see that some of its values are repeated. Hence array of IET values can be grouped by sets with the same values (for details see Chapter 4). As a result, we can see that the array of IET values has a layered nature in all described cases. The number of layers will increase after a hypervisor is loaded and this fact can also be used for hypervisor detection.

The revealed IET variability indexes, variance (or second moment) and number of layers (or spectral width) are resilient to time cheating. Hypervisor can only decrease the mean value of IET but not the variability characteristics.

As a first approximation this analysis reveals two theoretical hypervisor indicators. This result is based on a hypothesis but now has to be comprehensively verified by experiments.

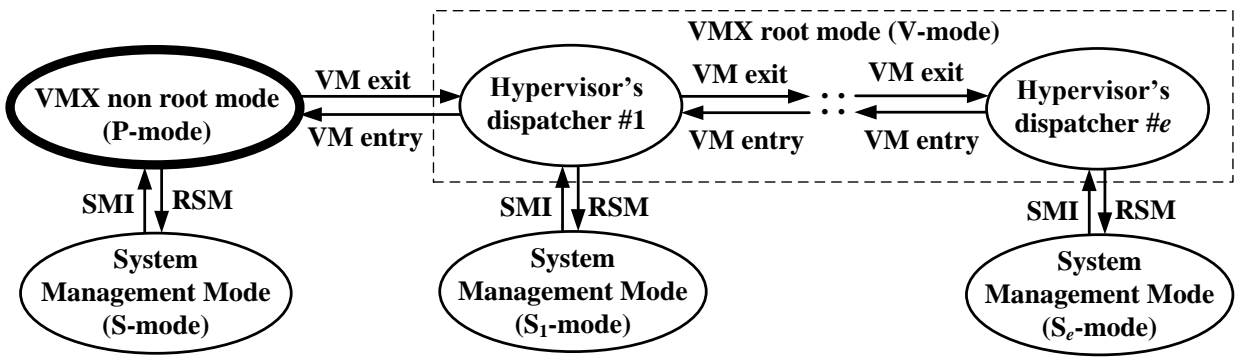

Figure 4 Switching between Modes with Several Nested Hypervisors

## 4. INSTRUCTION EXECUTION TIME RESEARCH & NEW STEALTH HYPERVISORS DETECTION ALGORITHMS

Probabilistic hypervisor detection is discussed in following papers (Desnos et al., 2011; Fritsch, 2008; Jian et al., 2010; Morabito, 2012). All these methods work only if a hypervisor is not hiding itself. What is more, these papers do not give enough attention to the random nature of IET.

Detection of stealthy hypervisors faces two challenges: time cheating and data fluctuations, which will be described in this paper.

### 4.1. Probabilistic Nature of Instruction Execution Time

Desnos, Filiol and Lefou (2011) suggested that the instruction execution time is normally distributed and there are no problems with precision (repeatability and reproducibility) of the measurement data.

However, all our experiments on different PCs showed that the measurement data are non-normally distributed. There are no well-known distribution patterns which these data would match. Moreover, data fluctuation is so large that mean and variance statistics differ significantly between sets of experiments. Therefore the precision of the measurement data does not comply with ISO 5725 (2004) requirements.

We have to take into consideration that outliers and jumps (discontinuity) are very common, which will alter statistical values, see Figure 5. A possible reason for outliers and jumps is the pipeline of instructions. Due to the fact that the time measurement procedure is quite simple and a PoC hypervisor with time cheating can be used, we can receive an abundance of experimental data for research and detection phase, which significantly helps. Relying on the probabilistic nature of IET we dealt with when setting up experiments, these revealed data peculiarities, processing of preliminary data, only appeared after that we applied statistical methods.

### 4.2. Experiments on Measurements of Instruction Execution Time

To detect a hypervisor we improve the detection method, which uses unconditionally intercepted instructions. We analyze IET sets in the two cases with a hypervisor and without any.

Experimental data was received by measuring a set of ten CPUID instructions by using RDTSC in a loop in Windows driver, see Figure 6. To dismiss the influence of other apps and drivers in the OS we ensured thread affinity with certain CPU core and raise IRQL to its maximum level. It is also possible to use deferred procedure call (DPC) to achieve an exclusive access to the hardware. An example of this scheme is described by Blunden (2012).

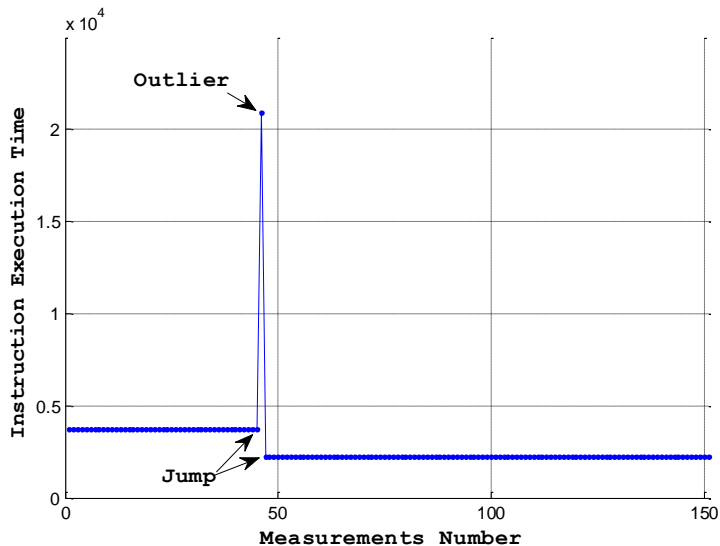

Figure 5 Scatter Plot of IET Array Fragment with One Outlier and Jump

We use CPUID instruction as an unconditionally intercepted one by any Intel-based hypervisor and also as a serializing instruction which prevents out-of-order execution (Intel, 1998; Fog, 2014).

Our proof-of-concept hypervisor (PoC hypervisor) is based on the VMM framework by Embleton (2007) with an added TSC cheating function. There are three different ways to cheat TSC: by TSC_OFFSET field in VMCS, catching execution of RDTSC or CPUID. We chose the last one: our hypervisor decreases the TSC's value every time CPUID is executed. This hypervisor's dispatcher is the smallest. By cheating TSC we can make sure that the average values of IET are the same to within one clock tick, whether the hypervisor is present or not. Therefore, this is the most complex case for detection.

To obtain data we used two nested loops. An example of an inner loop is shown on Figure 6, it was executed 1000 times without any delays. Outer loop was executed 10 times with a two-second delay between each iteration. The results of this experiment were recorded to a 1000x10 array (see Table 1); the columns contain data from inner loops.

According to ISO 5725 repeatability requirements we repeated the complete experiment five times with a two-second delay between each iteration. To control reproducibility of data we checked the results on 10 different days. All in all for this period we measured *50 arrays of 1000x10*, which will be further processed. That period was sufficient to reduce variation intervals of statistics: average values, variance etc.

Six PCs were involved in testing, see Table 2. In the first five PCs we used our PoC hypervisor, and in the last PC we used a specialized hypervisor loaded by BIOS – TRace EXplorer (TREX) by Tichonov and Avetisyan (2011).

```
KeSetSystemAffinityThread(affinity)
KfRaiseIrql(HIGH_LEVEL)
for (...)
{
    RDTSC
    MOV hst, EDX
    MOV lst, EAX
    CPUID // 1
    ...
    CPUID // 10
    RDTSC
    MOV hfin, EDX
    MOV lfin, EAX
    save_time(hst, lst, hfin, lfin)
}
```

Figure 6 Code Fragment for Obtaining Data

Table 1 Example of Array of Measured IET without a Hypervisor

| Measurement no | Inner loop iteration | | | |
|---|---|---|---|---|
| | **1** | **2** | **…** | **10** |
| 1 | 2896 | 2888 | … | 2896 |
| 2 | 2896 | 2888 | … | 2880 |
| … | … | … | … | … |
| 1000 | 2888 | 2888 | … | 2888 |
| Average of a column | 2895 | 2888 | … | 2888 |
| Variance of a column | 1738 | 1267 | … | 1196 |

Table 2 CPU Models and OS Versions

| PC# | CPU models and OS versions |
|-----|---------------------------|
| 1 | Intel Core 2 Duo E6300 / Windows 7 |
| 2 | Intel Core 2 Duo E8200 / Windows 7 |
| 3 | Intel Core 2 Duo E8600 / Windows Live CD XP |
| 4 | Intel Core i7 950 / Windows XP |
| 5 | Intel Xeon X5600 / Windows 7 |
| 6 | AMD Phenom X4 945 / Windows Live CD XP |

### 4.3. Peculiarities of Instruction Execution Time and Ways of Hypervisors Detection

Our experiments confirmed the following:

1. IET measured by TSC is a *random value*, which depends on a CPU model, OS version and on whether or not a hypervisor is present.
2. The average and variance of IET arrays are larger if a hypervisor is present than if it is not.
3. The difference between average and variance of IET arrays becomes more significant after every new nested hypervisor has been loaded.

We can easily and reliably detect a non-hidden hypervisor by just comparing the average values of IET arrays. The average values of IET arrays with a non-hidden hypervisor are almost 10 times larger than without it.

But a hypervisor can apply time cheating technique and as a result the average values of IET will be the same as corresponding values without a hypervisor. There are no time-based detection methods which work well under such circumstances. Our experiments were focused on this challenging case.

Using more common statistical methods in hypervisor detection proved to be inapplicable. The reasons will be given below.

By using statistics we can determine if there is a statistically significant difference between two sets of data. We already know which of the set will be measured with exposure and without it.

But in the current situation we have several sets. We can connect several sets to a big one, and use classical approaches, but such operation has to be proved. For this case there are no proven statistical methods.

Applying current approaches to determine significant difference between the sets did not yield any positive results for a variety of reasons. We consider the columns of arrays as a random sample, also as a result of the random process. It is impossible to use the first method, because of the fluctuation of measurements and lack of homogeneity. The second method is not applicable either, because of overlapping variation intervals and instability of characteristics.

We see that homogeneity of variances (HOV) is violated in all our experimental data, and as a result we cannot use analysis of variance (ANOVA) in data processing.

We conclude that methods of parametric and non-parametric statistics are not applicable in the current situation. That is why we developed the following methods, including the present author's approaches:

1. Low-frequency filtration.
2. Calculation of experimental probability.
3. Two-step way to calculate statistics.
4. Variation interval as confidence interval.
5. Iteration of measurements if the statistical value is at the intersection of two variation intervals.

Due to filtration we can decrease fluctuation and stabilize variation characteristics.

Due to calculation of experimental probability we can find threshold values and so minimize type I and II errors.

We choose a two-step way of calculating in order to reduce overlapping of these characteristic intervals.

To calculate a confidence interval we choose the idea of the confidence interval method of Strelen (2004) and Kornfeld (1965), in which a confidence interval is calculated as a variation interval or difference between maximum ($S_{max}$) and minimum ($S_{min}$) values of statistic. The confidence level is the following $P\{S_{min} \ll \theta < S_{max}\} = 1 - 0.5^{n-1}$, where '$n$' is the length of a sample.

We have to study a situation when a calculated statistical value will be at the intersection of two variation intervals. In this situation it is impossible to decide whether a hypervisor is present or not. In this case we have to repeat measurements of IET arrays and calculations of statistics. In accordance with the multiplication theorem of probability a recurrent hit in the intersection zone is unlikely.

### 4.3.1. Applying Layered Structure of IET Arrays to Hypervisors Detection.
Numerous experiments show that IET arrays have a layered structure. It means that each IET array is comprised of several layers, whose characteristics depend on CPU, OS and whether or not a hypervisor is present.

First of all our experiments confirm, that the number of layers with a hypervisor is larger than without a hypervisor.

To make it clear, the results of an experiment are given below. We measured IET arrays in two cases: without hypervisor and with it.

The right part of Figure 7 is a scatter plot of the IET array; each point corresponding to the measured duration of ten CPUID instructions. Experiment numbers are on the x-axis, while IET values are displayed on the y-axis.

Blue color corresponds to IET without a hypervisor, red color corresponds to IET with a hypervisor, which is applying time cheating. This technique leads to getting approximately the same mean value if hypervisor is present with the mean value without a hypervisor.

The left part of Figure 7 shows the corresponding frequency polygons or relative frequency chart. We can see that with a hypervisor the number of polygon points (or number of layers) is larger than without a hypervisor.

The similar nature of polygons was also noted by Morabito (2012). His observations show that the data is generally not normally distributed and skewed, long-tailed data with outliers is fairly common. Similar plots of IET array fragments are given in the paper by Fritsch (2008) in the part

"A.4 Empirical results" and by Li, Zhu, Zhou, & Wang (2011). However, the fact that layered structure could be used for hypervisor detection had not been mentioned.

If several hypervisors are present, the layering structure of IET arrays is still obvious. We measured IET arrays in four different cases: without hypervisor (black), with only own PoC hypervisor (green), with only Acronis hypervisor (blue) and with two nested hypervisors (red). The scatter plots of the corresponding IET arrays are shown on Figure 8.

To make it clear, the scatter plots are spaced vertically. We can see that without a hypervisor the plot consists of only one line with quite rare jumps. If PoC hypervisor is present, the corresponding plot has 2-3 layers with significant jumps. The situation is similar if only Acronis hypervisor is present. If two nested hypervisors are present we can see that the plot becomes a cloud of points, there are a lot of layers with low frequency.

The best way to reveal the number of layers is to use the frequency distribution of measured IET arrays. We calculate frequency distribution with each class for one value or without intervals of numbers. Number of layers equals the number of classes.

It is possible to detect a stealth hypervisor, which uses the Blue Chicken technique. Temporary self-uninstalling of this hypervisor originally occurs after 50-100 measurements of IET because hypervisor needs to recognize time-based detection. As a result we will see the changed nature of the scatter plot: the first 50-100 measurements will have a layered nature and the remaining portion of measurements will have just 1-2 layers because the hypervisor has already been uninstalled. This changing of the scatter plot will be repeated in the next columns; because they were measured with a two-second delay.

However, our experiments show that direct use of these indicators is problematic for two reasons. These characteristics are not always constant (they are unstable) and also variation ranges of these characteristics overlap each other whether hypervisor is present or not. Later we will discuss how to deal with it.

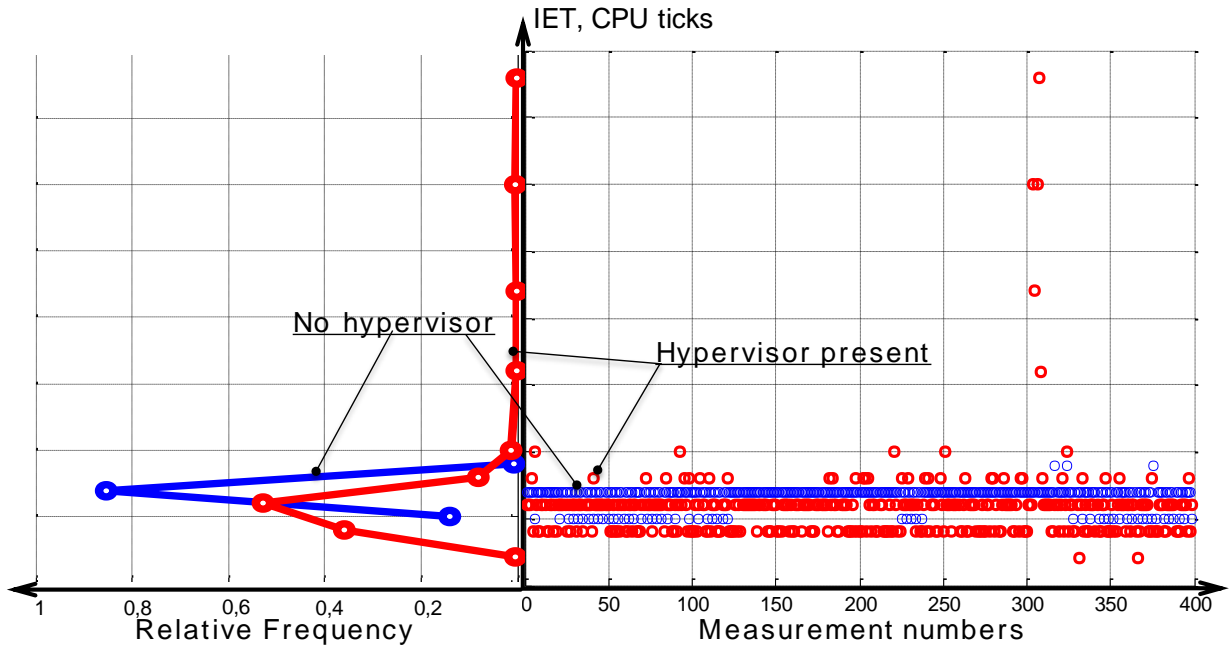

Figure 7 Scatter Plots of IET Arrays Fragments and Corresponding Frequency Polygons

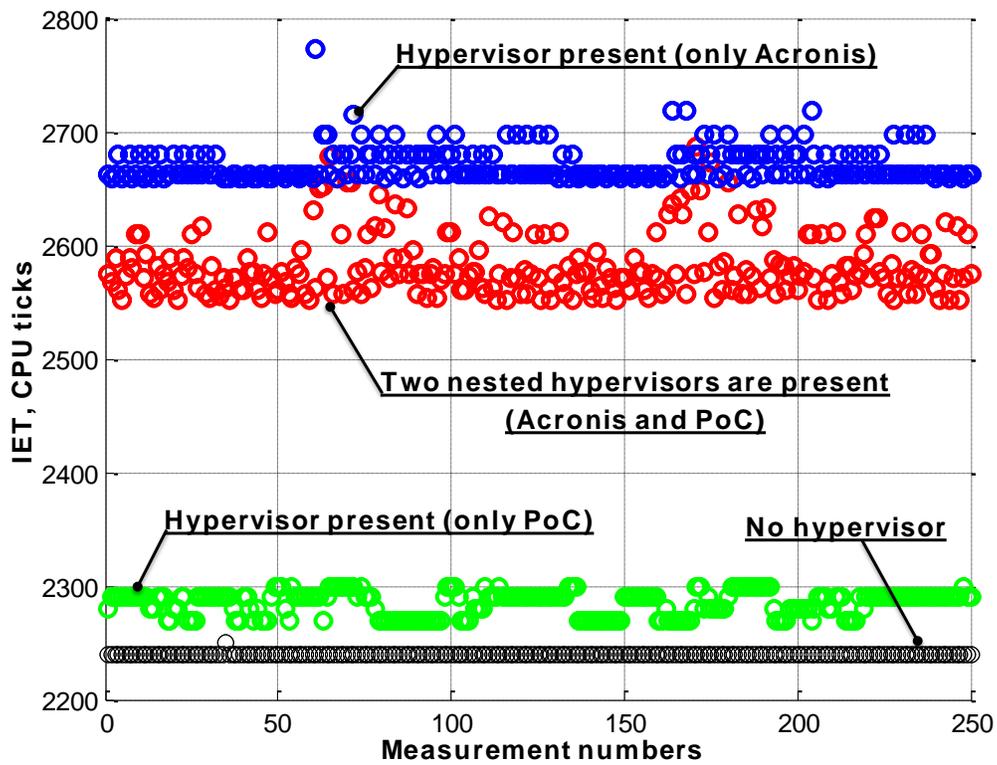

Figure 8 Scatter Plots of IET Arrays Fragments in Four Different Cases

**4.3.2. Applying Second and Fourth Moments to Hypervisors Detection.** All our experiments also confirmed the following result of section 3. After hypervisor is loaded the numerical values that measure the spread of IET arrays will increase. We obtained good results with the second and fourth moments. Moreover, after loading each nested hypervisor these sample characteristics increase, which is clearly seen in Figure 8. Experiments show that the sixth and higher moments of IET arrays are seriously inaccurate.

As mentioned before, outliers and jumps (discontinuity) significantly affected values of the second and fourth moments. That is why it is impossible to achieve a stable distinction between sample characteristics and draw a conclusion as to whether a hypervisor is present or not. Negative impacts of these factors can be eliminated by simultaneously applying two techniques: fitting or low-frequency filtering and *"length-averaging"*. We get sample characteristics before and after an outlier and calculate the final value by averaging of the corresponding fragments lengths for "length-averaging".

In order to reduce overlapping of these characteristics intervals we chose a two-step way of calculating. We calculate the second and fourth moments for each column in the table (IET array), see Table 1. This brings us to a set of these characteristics, which we consider as a new sample and repeatedly calculate characteristics of this set.

In other words, from the primary column of IET array we get the secondary characteristics, which we are processes by statistical methods. Consequently this helps us to significantly reduce or avoid the overlapping of new characteristics intervals.

All theoretical principles from Section 3 were successfully confirmed by experiment. The number of layers of IET arrays, second and fourth moments increased and remained on the same increased level after a new hypervisor was loaded, i.e. they can be used to detect a hypervisor and several nested ones. Moreover the ways of calculating threshold values of each statistic will be given with due consideration of data fluctuations.

## 4.4. How to Calculate Threshold Values of Statistics to Detect Hypervisors

Hypervisor detection includes comparison of calculated statistics values with threshold values. If statistical values are greater than threshold values, we conclude that a hypervisor is present, otherwise there is no hypervisor. The main goal is to find a suitable filtration level and the statistic, which has an appropriate threshold value or minimal sum of type I and type II errors.

To calculate threshold values we have to measure 50 arrays 1000x10 for two cases when a hypervisor is present or not, 100 arrays in total. We use own PoC hypervisor, because it contains the minimal set of instructions in CPUID dispatcher and its only role is TSC cheating. This is the most difficult case. PoC hypervisor's threshold values will help to detect any other hypervisor with more functions, as it will cause more changes to IET variation.

Calculating threshold values includes calculating statistics in two ways after low frequency filtering with the following levels {0, 0.02, 0.05, 0.1, 0.15, 0.2} or {0%, 2%, 5%, 10%, 15%, 20%}.

One way is to calculate statistics for each column 1000x1 of a 1000x10 array. After this calculation we analyzed the received set of 10 values as a new sample and then averaged them ($\bar{l}$ – "averaged columns value").

Another way is to calculate statistics for one big column 10,000x1 obtained from an array 1000x10 by vectorizing (Vectorization (mathematics), 2014) ($l_v$ – "vectorized array value").

It should not be forgotten that outliers and jumps (discontinuity) significantly change statistics values and therefore we have to delete them. We find a jump as the maximum value in the first order difference (The Scipy Community, 2009). The threshold value of a jump is 300 CPU ticks, which can then be corrected.

The calculation algorithm of threshold values is the same for all statistics and includes three steps:

1. Receive and process IET array every day. Receive preliminary results.
2. Process the preliminary results which are obtained for 10 days. Receive threshold values and probabilities of type I and II error.

3. Create the final table with all appropriate statistics.

We are going to describe a way to calculate threshold values of a new statistic – a number of layers.

The first step is to filter each column from Table 1 with different filtration levels. For each received column we calculate the number of layers. Calculated values are given in corresponding columns in Table 3.

The last but one column in Table 3 includes the mean values of the number of layers for each filtration level. For example the first value 12 is (28+29+...+10)/10. The last column includes the values of the number of layers, which were calculated from the column 10,000x1 for each filtration level. E.g. the first value 53 means the number of layers in the array 10,000x1 after its filtration with level 0%.

Table 3 Example of Calculating the Number of Layers If no Hypervisor is Present

| Filtering level | Values of the number of layers for each column in array 1000x10 | | | | Averaged columns value, $\bar{l}$ | Vectorized array value, $l_v$ |
|---|---|---|---|---|---|---|
| | 1 | 2 | … | 10 | | |
| 0 | 28 | 29 | … | 10 | 12 | 53 |
| 0.02 | 4 | 3 | … | 3 | 4 | 6 |
| 0.05 | 3 | 3 | … | 3 | 3 | 6 |
| **0.1** | **2** | **2** | **…** | **3** | **3** | **3** |
| 0.15 | 1 | 2 | … | 2 | 2 | 3 |
| 0.2 | 1 | 2 | … | 2 | 2 | 2 |

Table 4 Number of Layers of IET Arrays for 2 Cases when a Hypervisor is Present and not

| Code of experiments | No hypervisor | | Hypervisor is present | |
|---|---|---|---|---|
| | Averaged columns value, $\bar{l}$ | Vectorized array value, $l_v$ | Averaged columns value, $\bar{l}$ | Vectorized array value, $l_v$ |
| | 5 | 23 | 11 | 47 |
| | 4 | 18 | 11 | 52 |
| day #1 (Ig10) | 4 | 15 | 10 | 34 |
| | 5 | 21 | 13 | 53 |
| | 4 | 15 | 14 | 68 |
| | ... | … | … | … |
| | 4 | 20 | 19 | 102 |
| | 6 | 32 | 15 | 77 |
| day #10 (Ig19) | 6 | 32 | 16 | 79 |
| | 6 | 32 | 20 | 88 |
| | 10 | 50 | 21 | 105 |
| Variation intervals | [4, 14] | [10, 110] | [8, 21] | [29,105] |
| Threshold values | ≤ 7 | ≤ 32 | ≥ 8 | ≥ 33 |
| Type I error | 0.04 | 0.12 | – | – |
| Type II error | 0 | 0.16 | – | – |

We can see that with filtration level "0.1" the values of $\bar{l}$ and $l_v$ are stabilized, therefore we will use this filtration level for this PC in the future. The similar table is also created if PoC hypervisor is present. Four numbers, values $\bar{l}$ and $l_v$ in two cases when a hypervisor is present or not present, are evaluated from a single 1000x10 array in each case.

This procedure was repeated for each of five arrays 1000x10 every day, for 10 days.

After that we create a preliminary table with threshold values and type I and II errors, see Table 4.

Stabilization of statistics is obvious in both cases when a hypervisor is present and not. We managed to achieve this stabilization only due to filtration of jumps and *length-averaging*, as previously mentioned.

Variation intervals were determined according to minimum and maximum values of the statistics in the columns. Variation intervals overlap, therefore if statistical values get into this overlapping, it is impossible to reliably detect a hypervisor. In these cases we have to repeat IET array measurements.

We chose threshold values so that the sum of probability of type I and II errors is minimal. Type I error means that we conclude that the hypervisor *is* there according to calculations, while actually it *is not* there. The probability of a type I error is experimentally calculated as a number of values, which are greater than the threshold value. A type II error means that we conclude that the hypervisor is *not* there, while actually it *is* there. The probability of this error is also experimentally calculated as a number of values, which is smaller than the threshold value. In other words, we calculate the probability of type I and II errors with this formula $r/g$, where '$r$' is the number of values in the column, which are outside the threshold, $g = 50$ is the total number of values in the column. For detection we used only those statistics, whose sum of type I and II errors are less than 0.2 (or 20%).

Below is a fragment of the final table (Table 5) with all appropriate statistics for all tested PCs from Table 2.

$\bar{T}$ is the average value of IET from all arrays without a hypervisor and all other statistical notations are below in Table 6. As mentioned above we can calculate the statistics in two ways: for each column and after vectorization.

Our research findings suggest that threshold values depend on Windows version. For the same hardware threshold values for Windows XP and Windows 7 are different, variation intervals of statistics on Windows XP are smaller than on Windows 7. This occurs because Windows 7 enables more SMI handlers than Windows XP.

We performed similar experimental checks for nested hypervisors. We used the following iteration algorithm:

1. First, we obtained threshold values for the case without a hypervisor. To do this we measured IET arrays without a hypervisor and with our PoC hypervisor. We received that $L \leq 31$ (number of layers) means there is no hypervisor. The probability of a false positive is 0.14. $L \geq 32$ means a hypervisor is present. The probability of false negative is 0.06.

2. Secondly, we installed Acronis Disk Director, which loaded its own hypervisor. In the same way we obtained threshold values for this case. To do this we measured IET arrays with only the Acronis hypervisor and with two nested hypervisors: PoC and Acronis. We found out that $L \leq 67$ or more precisely $32 \leq L \leq 67$ means that only the Acronis hypervisor is present. $L \geq 86$ means that two nested hypervisors simultaneously work. Probabilities of type I and II errors in the latter case is 0.

Table 7 includes the threshold values for all mentioned cases.

Table 5 Final Table with all Appropriate Statistics

| PC | Statistics | Filtration level | Threshold values | | Probability | |
|---|---|---|---|---|---|---|
| | | | No hypervisor | Hypervisor is present | Type I error | Type II error |
| 1 | $\bar{T}$ | 0 | $\leq 2{,}911$ | – | – | – |
| | $\bar{L}$ | 0 | $\leq 7$ | $\geq 8$ | 0.04 | 0 |
| | $\bar{D}$ | 0 | $\leq 14$ | $\geq 18$ | 0.02 | 0 |
| | $\bar{M}$ | 0.1 | $\leq 679$ | $\geq 947$ | 0.02 | 0 |
| | $\mu$ | 0.1 | $\leq 104{,}161$ | $\geq 111{,}041$ | 0.02 | 0.10 |
| 2 | $\bar{T}$ | 0 | $\leq 2492$ | – | – | – |
| | $\bar{L}$ | 0 | $\leq 11$ | $\geq 12$ | 0.1 | 0.06 |
| | $\bar{D}$ | 0.2 | $\leq 100$ | $\geq 101$ | 0.08 | 0.1 |
| | $\bar{M}$ | 0.2 | $\leq 168$ | $\geq 13{,}030$ | 0.14 | 0.02 |
| 3 | $\bar{T}$ | 0 | $\leq 2{,}431$ | – | – | – |
| | $\bar{L}$ | 0 | $\leq 6$ | $\geq 8$ | 0 | 0 |
| | $\bar{D}$ | 0.1 | $\leq 15$ | $\geq 41$ | 0 | 0 |
| | $\mu$ | 0.1 | $\leq 609$ | $\geq 3{,}410$ | 0 | 0 |
| 4 | $\bar{T}$ | 0 | $\leq 5{,}018$ | – | – | – |
| | $\bar{L}$ | 0 | $\leq 22$ | $\geq 26$ | 0.02 | 0.02 |
| | $\bar{D}$ | 0.1 | $\leq 177$ | $\geq 181$ | 0.1 | 0.1 |
| 5 | $\bar{T}$ | 0 | $\leq 2{,}852$ | – | – | – |
| | $\bar{L}$ | 0 | $\leq 67$ | $\geq 71$ | 0.04 | 0 |
| | $\bar{D}$ | 0 | $\leq 16{,}416$ | $\geq 48{,}920$ | 0 | 0 |
| 6 | $\bar{T}$ | 0 | $\leq 2{,}126$ | – | – | – |
| | $\bar{L}$ | 0 | $\leq 34$ | $\geq 241$ | 0 | 0 |
| | $\bar{l}$ | 0 | $\leq 134$ | $\geq 593$ | 0 | 0 |
| | $\bar{D}$ | 0 | $\leq 216$ | $\geq 5{,}478$ | 0 | 0 |
| | $d$ | 0 | $\leq 345$ | $\geq 5{,}422$ | 0 | 0 |
| | $\bar{M}$ | 0.02 | $\leq 54$ | $\geq 956$ | 0 | 0 |

Table 6 Statistical Notations

| | Averaged columns value | Vectorized array value |
|---|---|---|
| Number of layers | $\bar{L}$ | $l$ |
| 2$^{\text{nd}}$ central moment | $\bar{D}$ | $d$ |
| 4$^{\text{th}}$ central moment | $\bar{M}$ | $\mu$ |

Table 7 Threshold Values for Two Nested Hypervisors

| **Threshold values** | **Conclusion about hypervisors and their numbers** | **Type I error** | **Type II error** |
|---|---|---|---|
| $L \leq 31$ | No hypervisor | 0.14 | 0 |
| $32 \leq L \leq 67$ | Only Acronis hypervisor is present | 0 | 0.06 |
| $L \geq 86$ | Two nested hypervisors are present | 0 | 0 |

**4.5. Detection of Stealthy Hypervisors**

According to experiments the detection of hypervisors goes in two stages: through preliminary and operational stages, see Table 8. First of all we have to make sure that there is no hypervisor in BIOS. To achieve this we update/flash BIOS with a known and trusted image. Malware in BIOS can prevent its updating by software utility. That is why the best way to overwrite BIOS is to desolder a microchip from the motherboard, flash it by hardware programmer and solder it back (Muchychka, 2013).

In the second step we install OS. We have to use official images to be certain that OS images do not include any malware or illegitimate hypervisors. Additionally OS components may be checked, e.g. by reverse-engineering.

In the third step we get threshold values by using PoC hypervisor. This step was described above.

In the fourth step we run the hypervisor presence check in an infinite loop. We measure IET arrays in a loop and compare calculated statistics with threshold values, which were calculated in step 3. We successively check if a hypervisor is present on each CPU physical core.

On the fifth and sixth steps we install supplementary software and monitor messages about new hypervisors.

If we get a message about new hypervisors after a program installation, we check if this hypervisor is legitimate. The approaches how to do this are beyond the scope of this paper. It may be noted that we can do it by calling corresponding support service etc. Once we conclude that the hypervisor is legitimate, we have to adapt the detection tool by obtaining new threshold values (step 3). If we conclude that the hypervisor is illegitimate, it must be removed from the system. In some cases this is solved by just uninstalling the previously installed program. However in more complicated cases we have to check all the system components including the BIOS image.

All source codes of getting threshold values, PoC hypervisor and detection tool are here (Korkin, 2014). The tool for getting threshold values consists of two parts: subsystem for IET arrays acquisition (C++) and subsystem for threshold values calculation (MATLAB). PoC hypervisor was developed using C++ and ASM, and it is compiled with Visual Studio. The detection tool consists of two parts: subsystem for IET arrays acquisition and subsystem for threshold values checks by MATLAB.

**5. CONCLUSIONS, DISCUSSIONS AND FUTURE WORK**

1.  Hypervisor detection is a well-known challenge. Malware with hypervisor facilities are serious information security threats. Many authors and companies are trying to tackle this challenge.
2.  In this paper we focused on and improved time based detection by unconditionally intercepted instructions. We studied the case when a hypervisor uses time cheating and temporary self-uninstalling to prevent its detection. For this situation appropriate time based approaches are not available on the Internet. Only the described methods are able to detect stealthy hypervisors in all known countermeasures: time-cheating etc.

Table 8 Detection of Stealthy Hypervisors

| Stages | Stage description |
|---|---|
| Preliminary | 1. Flash BIOS with a trusted image or firmware.<br>2. Install OS.<br>3. Get threshold values for no hypervisor case. |
| Operational (detection) | 4. Check in a loop if a hypervisor is present.<br>5. Install supplementary software (optional).<br>6. Monitor messages about a hypervisor presence.<br>7. To adapt the tool to new legitimate hypervisor go to 3. |

3. We explored the probability characteristics of instruction execution time and proposed a new technique for the detection of a hypervisor and several nested ones.
4. We developed techniques for calculating threshold values of various statistics and a step-by-step guide on how to detect a hypervisor.
5. These methods work well on different PCs with both Intel and AMD CPUs and detected PoC hypervisor and special BIOS hypervisor.

### 5.1. Hypervisor Detection without Flashing BIOS and Reinstalling OS

The proposed hypervisor detection method (or its preliminary starting procedure) needs to stop system activity to flash BIOS, reinstall OS etc. But for some systems this interruption of work is prohibited or impossible. However, on the basic of our experimental results, we can guarantee no hypervisor presence without performing 1-2 steps and unwanted system shutdown. To achieve this we acquire IET arrays on PC, which is already in operation. If after IET arrays filtering step we get 1-2 stable layers, this will mean that there is no hypervisor. This peculiarity occurs on PCs with Windows XP and should be investigated further.

### 5.2. Applying Numerical Values of Layers for Hypervisor Detection

We have discovered another pattern which can be used to detect a hypervisor. Thus most of our experiments numerical values of layers are unique. For example, in Figure 7 we see that numerical values of different layers after filtering indicate hypervisor presence. We achieve the following numerical values of layers without hypervisor {2160, 2168, 2184, 2192, 2200, 2478, 2480, 2880, 2888, 2904, 2920, 2936} and these values {2876, 2884, 2892, 2900, 2908, 2916, 2924} with PoC hypervisor. We see that these two sets do not contain equal values. Moreover, if a hypervisor cheats TSC so that the first members from each set are equal, the second and the next members from the above sets will differ. This happens because of the differences of deltas in each set {8, 24, 32, 40, 318, 320, 720, 728, 744, 760, 776} and {8, 16, 24, 32, 40, 48}.

The reasons for this difference and its resilience to hypervisor countermeasures requires further research.

### 5.3. Ways to Develop Computer Forensics and Statistics for Universities

The proposed statistical methods and tools for hypervisor detection can be used in two different disciplines. Firstly, it may become a part of Computer Forensics Discipline, when students can acquire practical skills working with hardware virtualization technology. PoC hypervisor can be used as a basic platform for further improvements; for example to create an event tracing tool which will monitor low-level events and will be resilient to modern malware. Hypervisor detection tools can be used to invent new detection approaches, based for example on all unconditionally intercepted instructions (CPUID, INVD, MOV from/to CR3, all VMX instructions, RDMSR and WRMSR). These may have different parameters, including wrong or invalid parameters, as well as profiling execution time for different sets and sequences of instructions, not just ten CPUIDs as is described in this paper. Analysis of time of physical memory access can be applied to find time anomalies due to possible hidden objects. Such a detection approach may need checking all the memory pages, including valid and invalid addresses. We compare IET characteristics before and after disabling the CPU's cache control mechanism. A stealth hypervisor has to cheat TSC with different deltas for each case, which does not always occur.

Secondly it may become a part of a course in "Statistics and Data Analysis". Because of its opportunity to acquire a lot of real experimental data sets students can acquire practical experience of data processing and its analysis. They can learn how to solve repeatability and reproducibility problems. They can apply different statistical criteria to test correlations between arrays for different cases: with a hypervisor and without it. As a result students will not only better understand the theoretical materials of the course, but will also acquire new practical skills and apply them in their own research.

### 5.4. Applying Hidden BIOS Hypervisor to Track Stolen Laptops

It is well known that an organization has to pay heavily every time an employee's laptop is lost or stolen. The idea is to create a software agent which will track a laptop, block it if it is stolen, control it remotely etc. This tool will work like Computrace LoJack by Absolute Software (2014). The key moment is to create a software agent, which will be really hard to detect, delete and block. By using hardware virtualization technology we can create a hypervisor, which works permanently. To guarantee that autorun works well, it will be loaded from BIOS. This hypervisor can hide memory areas and prevent its own rewriting by software tools with the help of Shadow Page Tables for AMD CPUs or Extended Page Tables for Intel CPUs. This hypervisor can be easily planted in any PC which supports hardware virtualization. To facilitate development of this hypervisor we can use open source software components, for example Coreboot (2014) for BIOS firmware, TianoCore (2014) for UEFI and XEN (The Xen Project, 2014) as a basis for this hypervisor.

### 5.5. Applying Hypervisor as USB Firewall to prevent BadUSB attack

Nohl and Lell (2014) presented an idea and prototype of malware USB stick. The idea lies in reprogramming a USB device in order to add new unauthorized functions. As a result, for example, a reprogrammed USB stick will work as a USB keyboard and by running malware commands can take over a computer. This vulnerability is really serious because this USB device works invisibly for user and AVs and formatting USB flash does not erase malware firmware.

We can solve this challenge by using a hypervisor's facilities, which will control all the devices access to the PC. By applying manual configuration mode the hypervisor can block malware activities of such devices. It will look as if a hypervisor is playing the role of a USB firewall. For example, after a USB device plugs into the computer port the hypervisor will display the list of all registered devices and allow the user to choose the appropriate position. After that the hypervisor will control the work of all USB devices according to the access policies of these devices. As a result this hypervisor working as USB firewall can guarantee protection of PCs from BadUSB attack or other malware USB devices.

## 6. ACKNOWLEDGEMENTS

I would like to thank Andrey Chechulin, Ph.D research fellow of Laboratory of Computer Security Problems of the St. Petersburg Institute for Informatics and Automation of the Russian Academy of Science (Scientific advisor – Prof. Igor Kotenko) for his insightful comments and feedback which helped us to uplift the quality of the paper substantially.

I would like to thank Iwan Nesterov, head of the department of mathematical methods of cyber security, LLC "CSS Security", Moscow, Russia for his invaluable contribution and support.

I would also like to thank Ben Stein, teacher of English, Kings College, London, UK for his invaluable corrections of the paper.

I am grateful to my grandfather Peter Prokoptsev, Ph.D, Kirovograd, Ukraine for his help with statistics and data analysis.

## 7. AUTHOR BIOGRAPHY

Igor Korkin, Ph.D has been in cyber security since 2009. He works at the Moscow Engineering & Physics Institute, training post-graduate students and supervising students. Research interests include: rootkits and anti-rootkits technologies. Took part at the CDFSL in 2014.